\documentclass[11pt,a4paper]{article}
\usepackage[margin=1in]{geometry}
\usepackage[T1]{fontenc}
\usepackage{lmodern}
\usepackage{amsmath,amssymb,amsfonts,braket}
\usepackage{array,booktabs}
\usepackage{subfig}
\usepackage{textcomp,url,verbatim,graphicx}
\usepackage[section]{placeins}
\usepackage{cite}
\usepackage{microtype}
\usepackage{hyperref}
\date{}

\begin{document}

\title{Image Compression Using Quantum Wavelet Transform and Quantum Convolutional Networks}

\author{Harshdeep Jadhav\thanks{College of Computing and Data Science, Nanyang Technological University} \and
Sreeraj Rajan Warrier\thanks{Department of Physics, Mahindra University, Hyderabad, Telangana 500043, India} \and
Jayasri Dontabhaktuni\footnotemark[2]}
\maketitle

\begin{abstract}
This paper presents a hybrid quantum-classical framework for grayscale image compression and decompression, leveraging the strengths of quantum computing and deep learning. The compression pipeline integrates a Variational Quantum Daubechies Wavelet Transform (V-QDWT) and a trainable Quantum Convolutional Neural Network (QCNN) optimized end-to-end to achieve efficient, image-adaptive multi-resolution analysis and entanglement-based feature reduction. Input images are encoded using the Normal Arbitrary Superposition State (NASS) representation, enabling compact and scalable quantum storage. For decompression, we implement inverse QCNN and V-QDWT circuits to reconstruct coarse image features natively, followed by a classical Super-Resolution Generative Adversarial Network (SRGAN) to enhance perceptual quality. Experimental evaluations on benchmark grayscale datasets demonstrate the efficacy of our hybrid approach. By jointly training the quantum layers, the base quantum pipeline closely rivals classical JPEG2000 standards. Subsequent SRGAN refinement substantially pushes the boundaries of the reconstruction, achieving superior structural fidelity (PSNR: 30.0667 dB, SSIM: 0.8744, Histogram Correlation: 0.9244). Histogram analysis and qualitative comparisons further validate the restoration of fine textures and intensity distributions. Our findings highlight the potential of combining variational quantum compression with classical deep learning to enable efficient, scalable, and perceptually-aware image processing in quantum-enhanced computing environments.
\end{abstract}


\section{Introduction}
With the ever-increasing amount of data being generated across various fields, the need for data compression is growing rapidly. It is a crucial area because large media files, such as images and videos, require compression to be stored efficiently on devices, and with increasing sizes of digital images, especially in the field of medical imaging, satellite imaging, robotics, etc., archiving and storing immense volumes of data presents significant technical challenges \cite{balasubramani2025quantum}. Classical computation for these massive images demands substantial memory and hardware resources while necessitating advanced data compression to optimize bandwidth utilization. As the application of image processing grows daily across domains such as robotics, medical diagnostics, defense, and remote sensing, dealing with bulky image data leads to increased algorithm complexity in classical processing systems. Therefore, image compression plays a vital role in minimizing storage footprint, accelerating transmission speeds, and reducing bandwidth overhead \cite{rohil2021insights}. This is particularly critical for resource-constrained edge devices with limited physical storage and low power budgets \cite{rohil2021insights}. The fundamental objective of image compression is to eliminate redundant and repetitive data from an image array while strictly preserving critical semantic information. This redundancy is conventionally classified into inter-pixel redundancy, coding redundancy, and psycho-visual redundancy. Consequently, standard compression frameworks comprise three significant phases: (i)~reduction of information redundancy, (ii)~reduction of spatial entropy, and (iii)~entropy encoding.

Compression techniques are broadly classified into two primary methodologies: lossless and lossy compression, differentiated by the exactness of detail preservation post-decompression.
Standard lossless compression methods, such as PNG and GIF, utilize entropy-reduction schemes such as Huffman coding and LZW coding \cite{rohil2021insights, vijayvargiya2013survey}. They typically encode all structural information. While lossless algorithms preserve absolute data fidelity, they inherently yield lower compression ratios.
On the other hand, lossy compression approaches achieve substantially higher compression ratios by selectively discarding high-frequency or perceptually negligible data from the image array. Common lossy compression standards include JPEG \cite{wallace1991jpeg}, which applies the Discrete Cosine Transform (DCT) to map spatial data into the frequency domain, and JPEG2000 \cite{skodras2001jpeg2000}, which implements the Discrete Wavelet Transform (DWT) for advanced multi-scale subband analysis \cite{antonini1992image}. 

While lossy compression achieves higher compression ratios, it typically reduces the fidelity of the reconstructed image.
Furthermore, early theoretical extensions of these transforms into the quantum domain have demonstrated potential computational advantages \cite{yuan2017quantum}. More recently, classical deep learning approaches, particularly deep convolutional autoencoders tracing back to early foundational bottleneck architectures \cite{hinton2006reducing}, have established new baseline benchmarks in end-to-end optimized lossy image compression \cite{balle2016end}. The lossy compression pipeline typically integrates spatial transformation, coefficient quantization (truncating bit-depth allocations for transformed matrices), and entropy coding to eliminate remaining statistical redundancies.
The advent of quantum information science has introduced highly effective paradigms for advanced image processing and high-dimensional compression. Quantum image compression leverages quantum mechanical principles such as state superposition and multi-qubit entanglement, to map pixel arrays into compact Hilbert spaces \cite{deb2024quantum}. Crucially, a quantum register can encode spatial sequences of length $n$ using polynomial resources $O(n)$, whereas classical architectures scale exponentially $O(n \times 2^n)$ to process equivalent combinatorial states \cite{deb2024quantum, nielsen2010quantum}. Concurrently, Quantum Machine Learning (QML) has provided robust theoretical framework for processing high-dimensional visual data on near-term quantum devices \cite{biamonte2017quantum, yan2016survey}. In all these methods it is essential to encode classical images into a quantum representation.

Early representation models include the Flexible Representation of Quantum Images (FRQI) proposed by Le et al. \cite{le2011flexible}, wherein computational basis states encode spatial coordinates while normalized probability amplitudes capture color or intensity values. Preparing an FRQI state from an uninitialized register requires a polynomial sequence of basic single-qubit Hadamard gates and controlled rotations \cite{le2011flexible}. To overcome measurement limitations in complex scenes, the Novel Enhanced Quantum Representation (NEQR) was introduced by Zhang et al. \cite{zhang2013neqr}. NEQR encodes basis states directly to represent bit-depth arrays, using separate entangled registers for spatial indices to enable highly precise grayscale retrieval. To accommodate non-square aspect ratios common in real-world sensors, the Improved Novel Enhanced Quantum Representation (INEQR) extended these mappings to arbitrary rectangular grids \cite{jiang2015quantum}. As the complexity of multi-channel data grew, further models such as the Multi-Channel Quantum Image (MCQI) \cite{sun2013multi} and Novel Quantum Representation of Color Images (NQRCI) \cite{sang2017novel} were introduced to natively process RGB channels and complex cryptography schemes \cite{bhowmik2016quantum}. Addressing the need for hyper-dense mapping, Li et al. \cite{li2014multidimensional} formulated the $n$-qubit Normal Arbitrary Superposition State (NASS) for multidimensional arrays, encoding $2^n$ distinct spatial pixels strictly within an $n$-qubit register. NASS parameterizes continuous pixel intensities as rotational phase angles, assigning orthogonal sub-states to distinct coordinate axes to facilitate highly dense spatial storage. Building upon the base NASS formulation, advanced variants such as NASSRP (incorporating relative phase parameters) and NASSTC (utilizing three distinct state components) were introduced to formalize lossless and lossy quantum compression bounds \cite{li2014multidimensional}.

Among data reduction operations, the Quantum Wavelet Transform (QWT) serves as the direct quantum mechanical equivalent to classical subband filtering. Classical wavelet decompositions are highly valued for isolating localized multi-scale structures, forming the backbone of modern compression codes. Similarly, QWT implementations decompose complex statevector amplitudes across discrete frequency resolutions using highly optimized logarithmic depth circuits \cite{Fijany1999}. Simultaneously, parameterized Quantum Convolutional Neural Networks (QCNNs) have emerged as powerful architectures for spatial pooling and latent extraction. QCNN image compression was found to be efficient with compression ratios of 4:1 \cite{bada2023detection}. In order to explore the efficacy of QWT and QCNN, a unified compression framework is proposed in this paper.
By systematically substituting quantum sub-circuits with traditional deep learning autoencoders and classical discrete transforms, we explicitly benchmark the feature preservation, computational trade-offs, and downstream reconstruction bounds of hybrid quantum image processing.

\section{Methodology}

\begin{figure}[!ht]
    \centering
    \includegraphics[width=\linewidth]{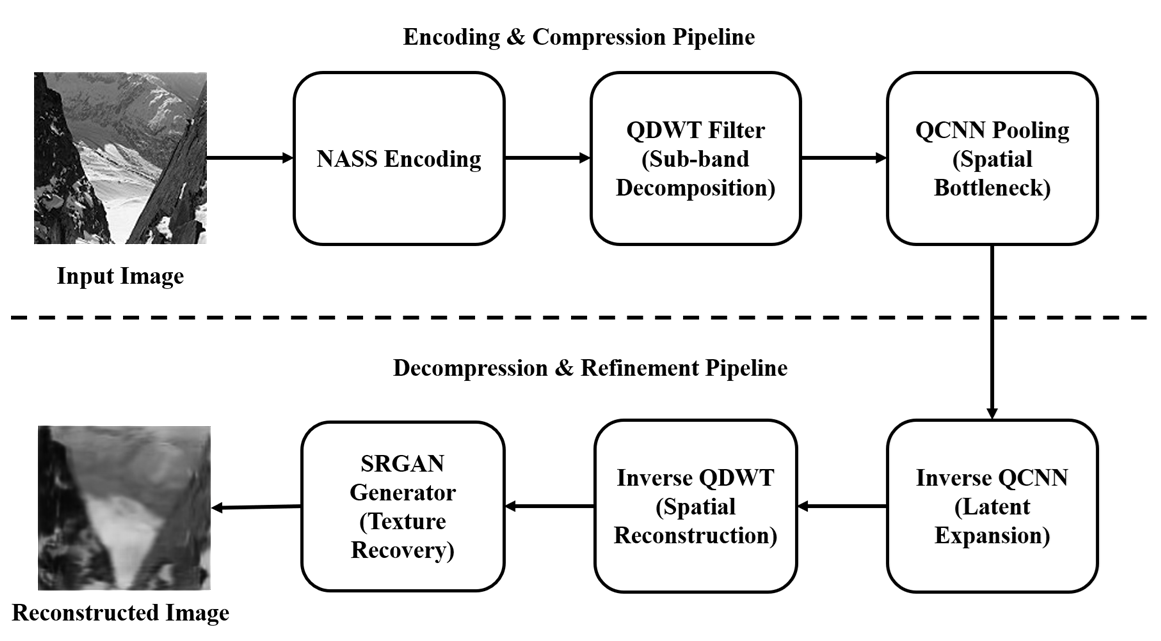}
    \caption{Overall schematic of the proposed hybrid quantum-classical image compression and decompression framework.}
    \label{fig:0}
\end{figure}

This section details our hybrid quantum-classical image compression and decompression pipeline. As illustrated in figure \ref{fig:0}, the proposed architecture seamlessly integrates state initialization, multiscale quantum subband filtering, feature bottle-necking, and classical generative refinement. First, a classical $2^n$-pixel image is encoded into a compact $n$-qubit statevector $\vert{}\Psi_\phi\rangle$ using the Normal Arbitrary Superposition State (NASS) mapping. The state vector undergoes multiscale spatial decomposition via the Quantum Daubechies Wavelet Transform (QDWT) to isolate frequency subbands. Next, a Quantum Convolutional Neural Network (QCNN) executes entanglement-based spatial pooling to compress the register into an ultra-compact latent state. During decompression, the latent state is expanded using inverse QCNN circuits and processed via the Inverse QDWT (IQDWT) to retrieve coarse image boundaries. Finally, a classical Super-Resolution Generative Adversarial Network (SRGAN) post-processes the reconstructed image to restore high-frequency textures and luminance distributions.

\subsection{NASS Representation}
To map classical image arrays into compact state-vectors, we implement the Normal Arbitrary Superposition State (NASS) framework proposed by Li et al. \cite{li2014multidimensional}. NASS utilizes $n$ qubits to simultaneously capture the spatial coordinates and intensity values of an image containing $2^n$ pixels. 

In this method pixel values are mapped to rotational phase parameters using specific functional transformations. For a standard 8-bit grayscale image $M=256$, where $M$ represents the total number of discrete intensity pixel values, the primary color-to-angle mapping $F_1$ is defined as:
\begin{equation}
    \label{eq:1}
    F_1: \text{Color} \longleftrightarrow \phi, \quad \phi_i = \frac{i \cdot \pi}{2(M-1)},
\end{equation}
where $i \in \{0, 1, \dots, 255\}$ corresponds to the scalar pixel intensity. Spatial positions are formalized over a $k$-dimensional Euclidean space $V$ spanned by orthogonal computational basis vectors, mapping local spatial coordinates to specific intensity values via $f: V \to \mathbb{R}$.

\begin{figure}[!t]
\centering
\includegraphics[width=\linewidth]{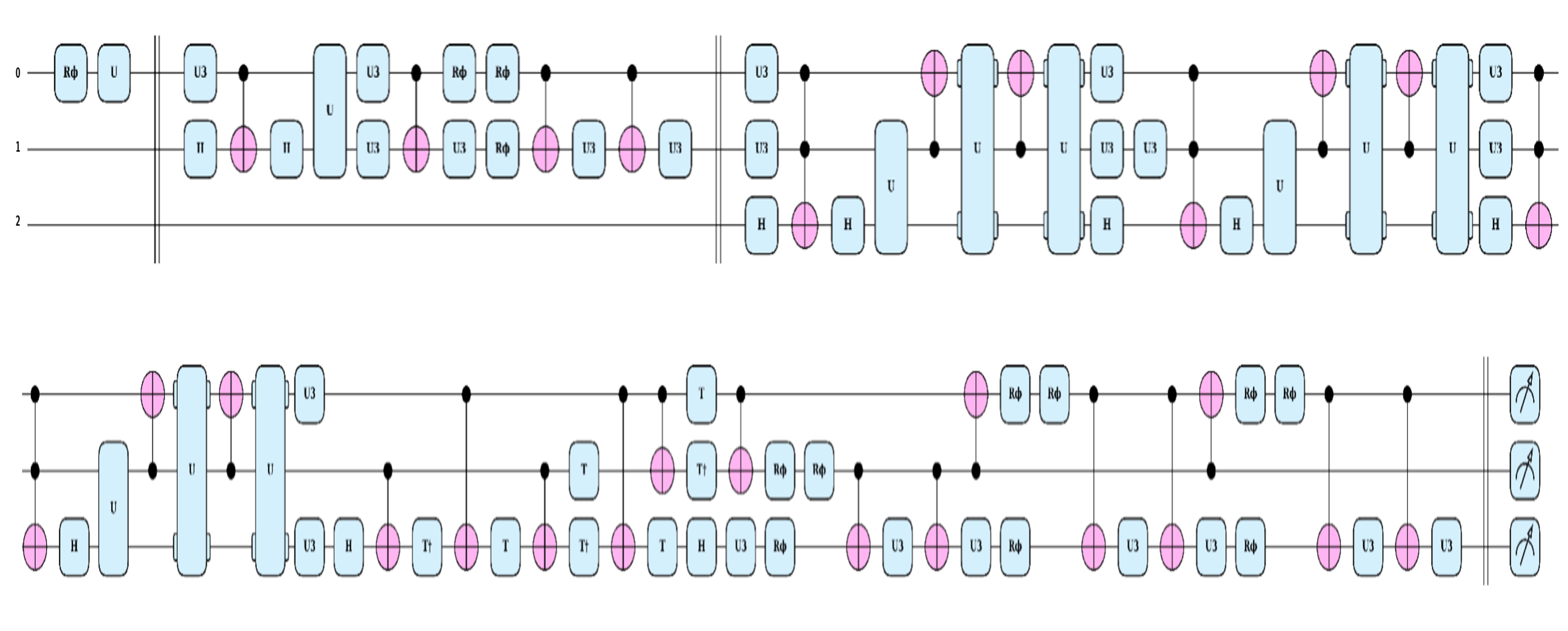}
\caption{Quantum circuit structure for preparing a NASS state.}
\label{fig:1}
\end{figure}

To accommodate secondary structural metadata or localized phase parameters, an auxiliary mapping $F_2$ is utilized, where \text{Number} refers to the discrete integer index of the auxiliary data being encoded \cite{li2014multidimensional}. This function assigns distinct rotational constraints, $\beta$—which mathematically bound the auxiliary phase parameters within specific continuous intervals to prevent interference with the primary intensity mapping—over these index sequences \cite{li2014multidimensional}. The unnormalized superposition state capturing the entire spatial field is formulated as:
\begin{equation}
    \ket{\Psi_{\phi}} = \sum_{i = 0}^{2^n - 1} a_i \ket{v_1}\ket{v_2}\dots\ket{v_k},
\end{equation}
where $a_i$ denotes the raw scalar pixel intensity, and $\ket{v_1\dots v_k}$ captures the binary spatial expansion indexing the targeted coordinate in Hilbert space. Because the underlying k-dimensional computational basis vectors are natively orthogonal, applying a standard Euclidean $L_2$ normalization yields the valid NASS encoding $\ket{\Psi_A}$:
\begin{equation}
    \ket{\Psi_A} = \sum_{i = 0}^{2^n-1} \theta_i \ket{v_1}\ket{v_2}\dots\ket{v_k}, \quad \theta_i = \frac{a_i}{\sqrt{\sum_{y = 0}^{2^n - 1} a_y^2}}.
    \label{eq:2}
\end{equation}
Incorporating auxiliary phase components extends the state representation to:
\begin{equation}
    \ket{\Psi_A} = \sum_{i = 0}^{2^n - 1} \theta_i \ket{v_1}\dots\ket{v_k} \ket{\chi_j},
    \label{eq:3}
\end{equation}
where $\ket{\chi_j} = \cos{\gamma_j}\ket{0} + e^{i\lambda_j}\sin{\gamma_j}\ket{1}$ isolates localized angular metadata.

Figure~\ref{fig:1} illustrates the specialized circuit architecture engineered to load a classical vector of size $2^n$ into the normalized probability amplitudes of an $n$-qubit register. The state preparation algorithm executes a hierarchical sequence of multi-controlled $R_y$ rotations from coarse subbands down to fine spatial details. At each step of the spatial hierarchy, the target rotation angle is computed recursively via vector norms:
\begin{equation}
    \theta = 2 \arctan\left(\frac{\| \mathbf{a}_\text{detail} \|}{\| \mathbf{a}_\text{approx} \|}\right),
\end{equation}
where $\| \mathbf{a}_\text{approx} \|$ and $\| \mathbf{a}_\text{detail} \|$ represent the Euclidean norms of the leading and trailing halves of the localized target array. The resulting controlled rotations are synthesized into standard single-qubit operations combined with optimized multi-controlled X (MCX) gates without utilizing ancillary helper qubits, ensuring highly dense spatial mapping suitable for highly-entangled, multi-layered quantum subband filtering.

\subsection{Quantum Daubechies Wavelet Transform}
Quantum Wavelet Transforms (QWTs) execute unitary spatial decompositions across amplitude-encoded registers, isolating high-frequency boundary textures from smooth structural approximations \cite{Fijany1999, Hirvensalo2004}. By implementing a parameterized Daubechies D4 wavelet kernel \cite{daubechies1992ten}, which is characterized by compact spatial support and high vanishing moments, the transform processes an $N$-pixel image array in parallel using logarithmic depth circuits $O(\text{poly}(\log N))$, offering substantial speedups over classical matrices.

\begin{figure}[!ht]
    \centering
    \includegraphics[width=\linewidth]{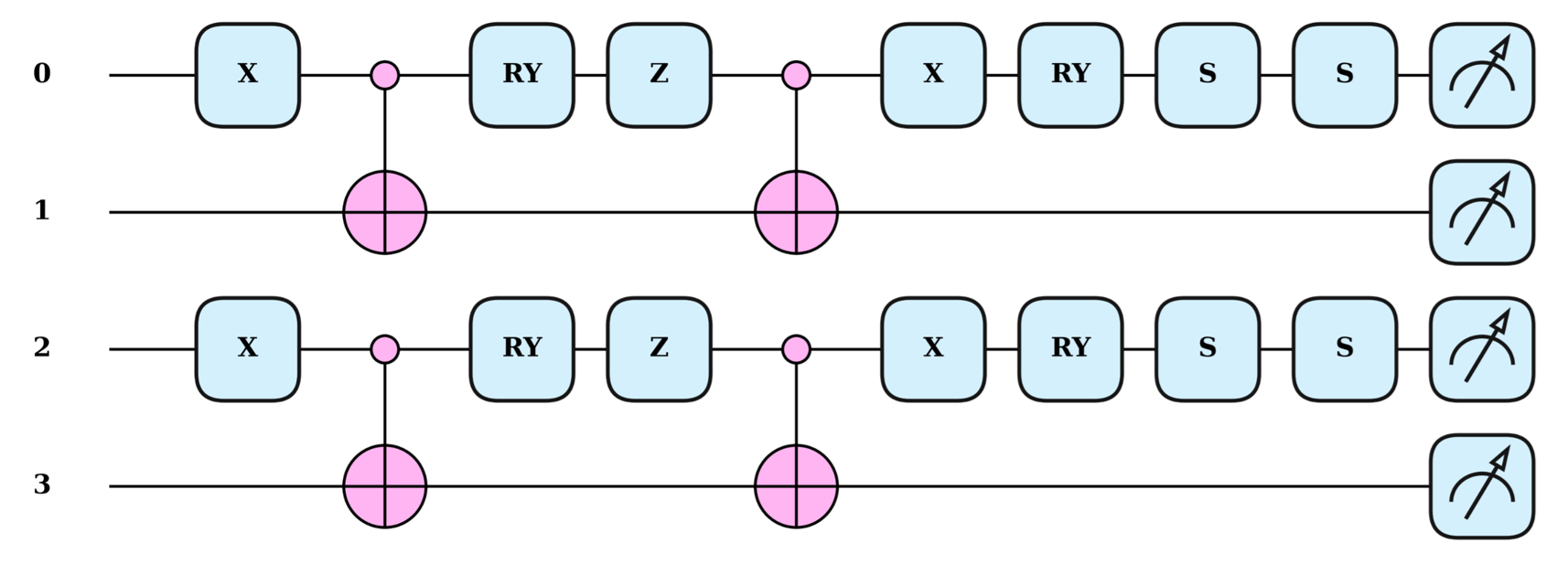}
    \caption{Quantum circuit implementing a 4-qubit Daubechies-like wavelet transform using multi-controlled gates and rotation-based filtering.}
    \label{fig:2}
\end{figure}

The forward Variational Quantum Daubechies Wavelet Transform (V-QDWT) applies local parameterized filter operations across adjacent probability amplitudes. The total unitary mapping $U_{D4}$ maps the input state vector to its multi-scale subband representation:
\begin{equation}
    \ket{\Psi_W} = U_{D4} \ket{\Psi_A} = \sum_{u,v} W[u,v] \ket{u,v}.
\end{equation}

To maximize spatial feature retention, the fixed-rotation amplitude filters traditionally used in QWTs are replaced with trainable parameterized quantum gates. This variational approach enables the wavelet kernels to dynamically adapt to the structural and frequency distributions of the input image. Structurally, $U_{D4}$ executes these trainable local filter operations ($S_0$ and $S_1$) combined with multi-controlled permutation cascades that reorganize sub-state amplitudes into discrete Low-Low (LL), Low-High (LH), High-Low (HL), and High-High (HH) spatial frequency bands.

Figure~\ref{fig:2} details a 4-qubit Daubechies-like wavelet decomposition. The circuit executes single-qubit trainable rotation operations followed by multi-controlled X cascades, conditionally propagating local phase shifts across the register to simulate classical filters.

\begin{figure}[ht]
    \centering
    \includegraphics[width=\linewidth]{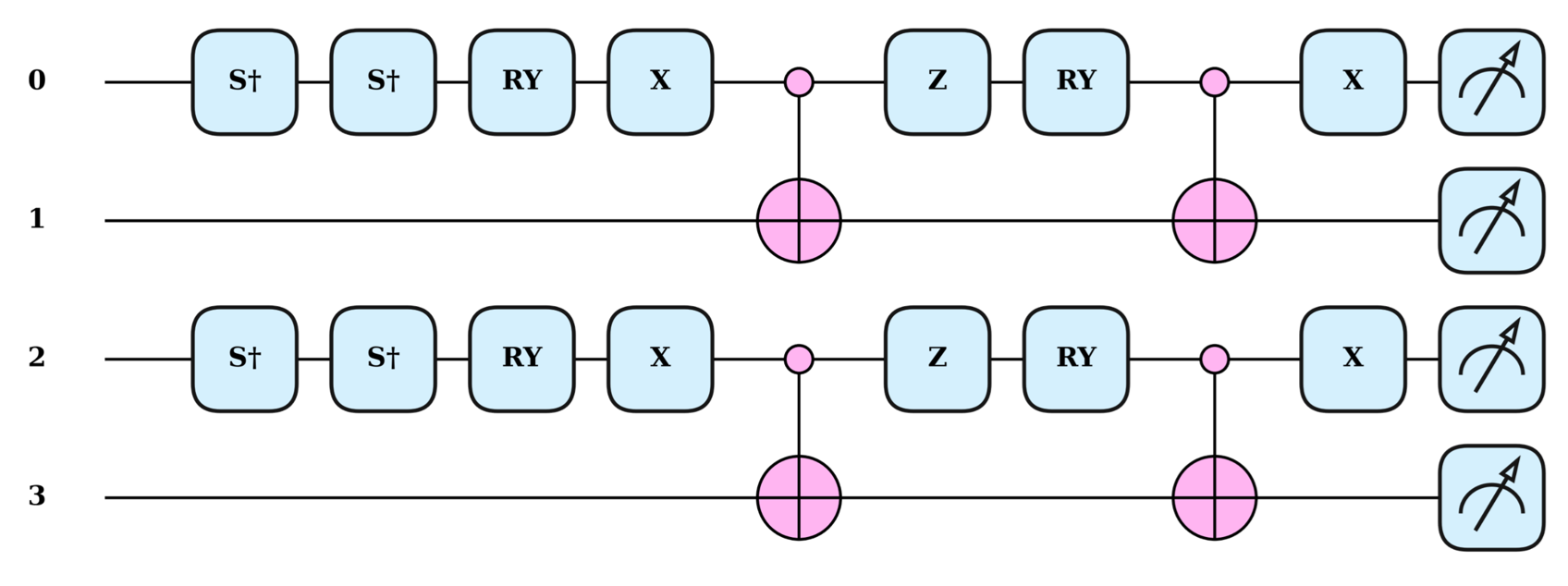}
    \caption{Quantum circuit implementing the inverse Daubechies D4 wavelet transform on a 4-qubit register.}
    \label{fig:3}
\end{figure}

The Inverse Quantum Daubechies Wavelet Transform (IQDWT) applies the exact Hermitian adjoint $U_{D4}^\dagger$ to reconstruct original spatial pixel arrays from the isolated subband states:
\begin{equation}
    \ket{\Psi_I} = U_{D4}^\dagger \ket{\Psi_W}.
\end{equation}
The decompression circuit perfectly mirrors the structural topology of the forward pass in reverse execution order. Local filter operations are inverted by applying the Hermitian conjugate of the optimized rotational angles ($S_0^\dagger$ and $S_1^\dagger$), while reversed permutation trees merge detail subbands back into coherent spatial fields as illustrated in Figure~\ref{fig:3}.

\subsection{Quantum Convolutional Neural Networks}

\begin{figure}[!ht]
    \centering
    \includegraphics[width=\linewidth]{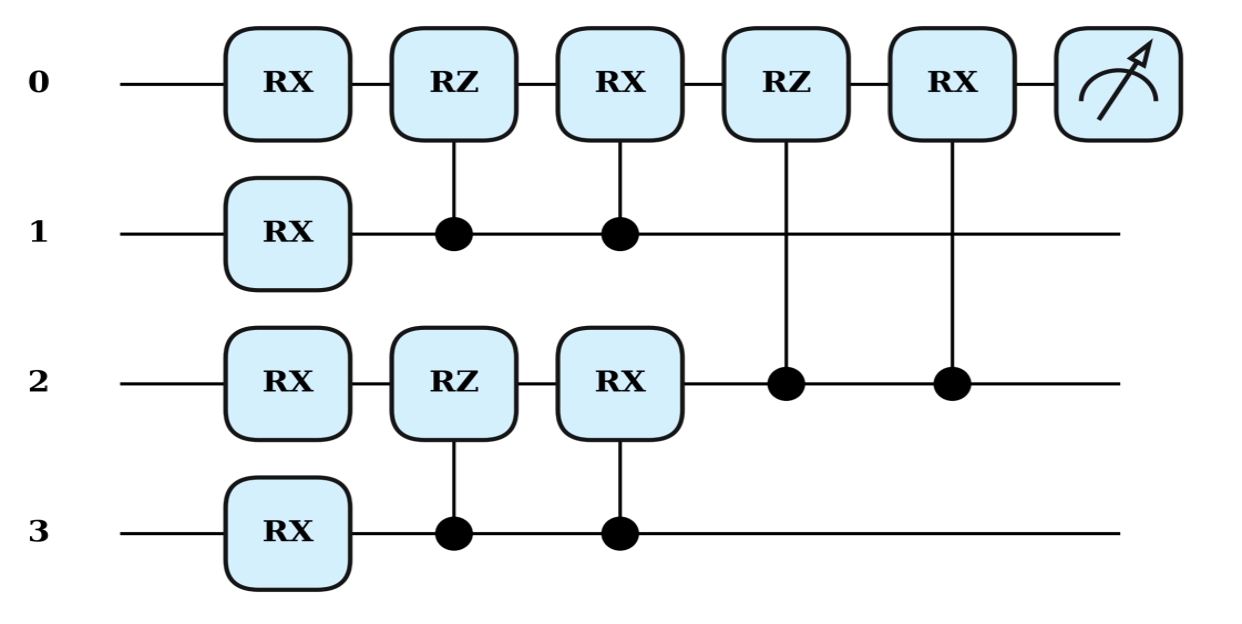}
    \caption{Quantum circuit used in QCNN for the Image Compression.}
    \label{fig:4}
\end{figure}

Following spatial filtering via the V-QDWT, the subband-encoded statevector undergoes spatial reduction using a Trainable Quantum Convolutional Neural Network (QCNN) \cite{Cong2019, Henderson2020}. 
The architecture is deliberately sequenced to apply the QCNN strictly after the V-QDWT, rather than processing raw spatial pixels. By first utilizing the V-QDWT to execute a multiscale decomposition, the framework successfully isolates high-frequency boundary textures from smooth structural approximations. Consequently, the downstream QCNN operates directly on these mathematically isolated frequency subbands, enabling highly efficient, entanglement-based spatial pooling and feature reduction without destroying critical edge structures during the quantum bottleneck.
It is known that parameterized quantum circuits (PQCs) and variational quantum algorithms (VQAs) are central to executing machine learning tasks on noisy intermediate-scale quantum (NISQ) hardware \cite{cerezo2021variational}. The forward QCNN ansatz implemented in this work utilizes highly entangled parameterized rotation gates (specifically $CR_Z$ and $CR_X$ operations) to process the localized $2 \times 2$ spatial patches within the 4-qubit register. Unlike static topological architectures that rely on fixed rotational parameters to avoid barren plateaus \cite{mcclean2018barren}, our framework actively leverages these parameterized operations as dynamically trainable pooling filters. The V-QDWT and QCNN modules are jointly optimized end-to-end as a Variational Quantum Autoencoder (VQAE). By training these parameters via classical backpropagation against a structural Mean-Squared-Error (MSE) loss function, the quantum network actively learns the optimal topological rotations to preserve critical spatial features natively within the computational basis. This learned feature retention drastically reduces the generative hallucination burden placed on downstream classical refinement.

It is known that parameterized quantum circuits (PQCs) and variational quantum algorithms (VQAs) are central to executing machine learning tasks on noisy intermediate-scale quantum (NISQ) hardware \cite{cerezo2021variational}. The forward QCNN ansatz implemented in this work utilizes highly entangled parameterized rotation gates (specifically $CR_Z$ and $CR_X$ operations) to process the localized $2 \times 2$ spatial patches within the 4-qubit register. Unlike static topological architectures that rely on fixed rotational parameters to avoid barren plateaus \cite{mcclean2018barren}, our framework actively leverages these parameterized operations as dynamically trainable pooling filters. The V-QDWT and QCNN modules are jointly optimized end-to-end as a Variational Quantum Autoencoder (VQAE). By training these parameters via classical backpropagation against a structural Mean-Squared-Error (MSE) loss function, the quantum network actively learns the optimal topological rotations to preserve critical spatial features natively within the computational basis. This learned feature retention drastically reduces the generative "hallucination" burden placed on downstream classical refinement.

To ensure actual data reduction in physical memory, the continuous Pauli-Z expectation values $\langle \sigma_z \rangle \in [-1, 1]$ extracted from the quantum bottleneck are uniformly quantized into 8-bit discrete integers. This classical quantization step maps the floating-point measurements to standard 8-bit depth prior to storage or transmission, guaranteeing that the 4:1 spatial dimensionality reduction achieved by the QCNN translates directly into a 4:1 reduction in physical payload. To achieve the final bit-rate, these quantized integers can be passed through standard lossless entropy encoders (e.g., Huffman or arithmetic coding) to eliminate remaining statistical redundancies, mirroring the classical JPEG2000 pipeline. During decompression, the encoded bitstream is decoded and symmetrically de-quantized back into continuous phase parameters to initialize the inverse quantum circuits.

\begin{figure}[!ht]
    \centering
    \includegraphics[width=\linewidth]{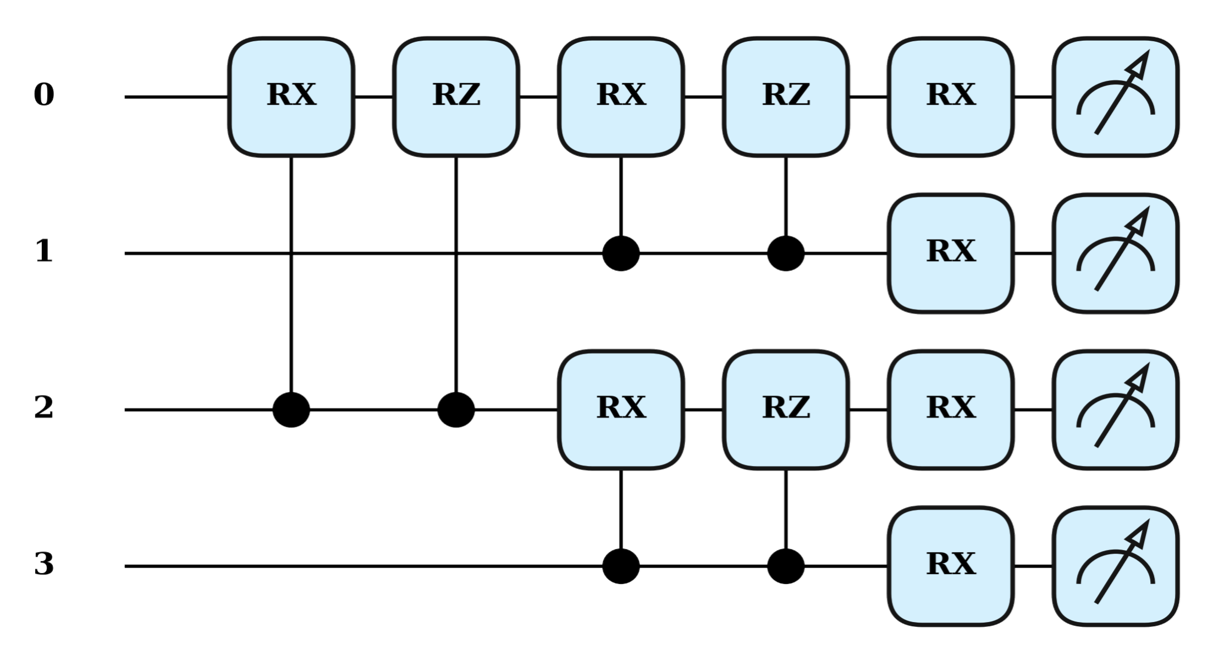}
    \caption{Quantum circuit implementing the inverse QCNN operation.}
    \label{fig:5}
\end{figure}

By ensuring the bit-depth remains constant (8 bits per value) before and after compression, the physical Data Compression Ratio (CR) is strictly defined by the spatial reduction induced by the measurement pooling:
\begin{equation}
    CR = \frac{N_{\text{original}}}{N_{\text{latent}}} = \frac{128 \times 128}{64 \times 64} = 4:1,
\end{equation}
where $N_{\text{original}}$ is the total number of pixels in the classical source image and $N_{\text{latent}}$ is the total number of measured Pauli-Z expectation values extracted from the quantum bottleneck. This provides a direct, mathematically equivalent baseline for comparison against the 3.97:1 physical compression ratio of the classical JPEG standard.

To restore spatial dimensions during decompression, the Inverse QCNN circuit applies the exact reversed sequence of parameterized operations utilizing the Hermitian conjugate of the dynamically optimized weights, as shown in Figure~\ref{fig:5}. This inverse framework maps compressed states back to unmeasured ancillary sub-registers, recovering full-dimensional probability amplitudes corresponding to reconstructed spatial subbands before feeding them into the inverse wavelet transform.

\subsection{Super-Resolution Generative Adversarial Network}
The image compression implemented using the proposed framework is inherently lossy in nature. To finalize the reconstruction, the Super-Resolution Generative Adversarial Network (SRGAN) is introduced exclusively at the final post-processing stage of the decompression pipeline \cite{Ledig_2017_CVPR}. Because the quantum bottleneck is now variationally trained to retain structural integrity, the base quantum decompression natively recovers a highly accurate coarse image. The SRGAN is then applied to bridge the final high-frequency structural gap without needing to heavily hallucinate missing macro-structures.
SRGAN was explicitly selected for this framework over standard CNN-based upsampling, such as SRCNN \cite{dong2015image}, because strict Mean-Squared-Error (MSE) optimization traditionally produces unacceptable blurring along high-frequency edges. Furthermore, while heavier generative architectures like ESRGAN exist \cite{wang2018esrgan}, the baseline SRGAN provides the optimal balance for this hybrid pipeline; its joint perceptual loss mechanism hallucinates and restores photorealistic semantic boundaries without introducing extreme computational overhead.
By building upon the foundational Generative Adversarial Network (GAN) framework \cite{goodfellow2014generative}, a methodology later expanded into models like ESRGAN, SRGAN prioritizes perceptual realism over pixel-wise averages. The SRGAN refinement pipeline utilizes a highly optimized Generator network $G$ composed of 16 deep residual blocks paired with efficient sub-pixel convolution upsampling layers (PixelShuffle) \cite{he2016deep, shi2016real}. The Generator maps the base quantum reconstructions $I^{LR}$ to sharp, photorealistic outputs $I^{SR} = G(I^{LR})$. The adversarial Discriminator network $D$ processes incoming images via a sequence of stride-2 feature convolutions to distinguish synthesized arrays from uncompressed ground truth distributions $I^{HR}$.

Network optimization is governed by a joint perceptual loss metric combining localized feature content mapping with adversarial feedback:
\begin{equation}
    \label{eq:4}
    \mathcal{L}_{\text{perceptual}} = \mathcal{L}_{\text{content}}(I^{SR}, I^{HR}) + \lambda \mathcal{L}_{\text{adversarial}}(I^{SR}).
\end{equation}

The feature Content Loss evaluates the Mean Squared Error across internal representation layers extracted from a pre-trained VGG19 deep network \cite{simonyan2014very} (specifically targeting the $\text{relu5\_4}$ activation volume). This specific perceptual loss formulation \cite{johnson2016perceptual} ensures the network prioritizes perceptual fidelity and texture recovery over strict pixel-wise mathematical averages:
\begin{equation}
    \mathcal{L}_{\text{content}} = \|\Phi(I^{SR}) - \Phi(I^{HR})\|_2^2.
\end{equation}

Simultaneously, the Adversarial Loss drives the creation of high-frequency textures by penalizing synthesized artifacts detected by the Discriminator:
\begin{equation}
    \mathcal{L}_{\text{adv}} = -\log D(I^{SR}).
\end{equation}
This joint learning mechanism ensures that the final decompressed frames recover absolute structural boundaries while generating natural image distributions. 

\subsection{Proposed Framework and Ablation Configurations}
To systematically establish the empirical contributions of the core quantum modules within our compression pipeline, we evaluate the primary full-quantum architecture using two targeted ablation configurations.

\paragraph{\textbf{Configuration A (Full Hybrid Pipeline)}}
The proposed architecture implements the complete end-to-end quantum-classical pipeline utilizing a highly parallelizable patch-based processing strategy. Input classical images are first resized to a standard $128 \times 128$ resolution. To operate within the strict coherence and gate-depth limits of current NISQ hardware, the global spatial grid is partitioned into localized $4 \times 4$ pixel patches. Each 16-pixel patch perfectly satisfies the $2^n$ dimensional requirement of the NASS framework and is mapped seamlessly into the probability amplitudes of a compact 4-qubit statevector without requiring boundary padding. Each 4-qubit register then undergoes multiscale subband filtering using the forward QDWT circuit. Subsequently, the forward QCNN applies entangling layers paired with deterministic measurement pooling to compress the 4-qubit state into a single latent expectation value. Tiling this operational cascade across all spatial patches reduces the image to a highly dense latent representation. During decompression, the unmeasured latent arrays are expanded using the Inverse QCNN circuit, processed via the IQDWT to recover spatial configurations, and measured classically to yield the coarse image reconstruction, which is ultimately finalized via SRGAN refinement.

\paragraph{\textbf{Configuration B (Ablating the QCNN)}}
To isolate the operational efficacy of quantum spatial pooling, Configuration B ablates the forward and inverse QCNN circuits, substituting them with a classical deep Convolutional Neural Network autoencoder. The classical input image is mapped to a quantum statevector via NASS and decomposed into subbands using the forward QDWT. The resulting probability amplitudes are measured and transferred to classical memory as a 2D feature array. This spatial matrix is routed through a classical deep CNN Encoder terminating in a Hyperbolic Tangent ($\text{Tanh}$) activation layer. The $\text{Tanh}$ activation strictly bounds the classical features within the continuous interval $[-1, 1]$, mathematically replicating the physical constraints of Pauli-Z expectation measurements utilized in the quantum pooling baseline. During decompression, a symmetric classical deep CNN Decoder reconstructs the subband coefficient array. The reconstructed array is directly fed into the quantum simulator via state preparation, bypassing standard NASS mapping to preserve local phase signs. Finally, the IQDWT circuit executes spatial reconstruction, and the classical output array is finalized via SRGAN processing.

\paragraph{\textbf{Configuration C (Ablating the QDWT)}}
To evaluate the multiscale filtering performance of the quantum subband transform, Configuration C ablates the forward and inverse QDWT circuits, replacing them with a classical baseline Discrete Wavelet Transform (DWT). The classical source image undergoes a 2D Daubechies D4 spatial decomposition executed classically using symmetric periodization padding to guarantee that subband coefficients map flawlessly into required binary array lengths without introducing spatial boundary overhead. Because classical high-frequency detail coefficients naturally contain negative scalar values that would be irreversibly destroyed by standard positive norm-based NASS preparation, the resulting DWT arrays undergo global Min-Max scaling to map all subband features into a strictly positive probability interval. The original minimum and maximum scaling parameters are cached as external metadata. The positive-shifted array is loaded into a quantum register via NASS encoding and compressed down to a bottleneck state using deterministic QCNN pooling layers. For decompression, the latent state is expanded using the Inverse QCNN circuit. Upon measurement, the cached scaling parameters are applied to mathematically shift the reconstructed subband array back into its native negative spatial spectrum. A classical Inverse Discrete Wavelet Transform (IDWT) reconstructs the spatial pixel frame, which is routed to the SRGAN network for final perceptual enhancement.

\section{Results}
The experimental validation of the primary hybrid compression framework and both the comparative ablation configurations was conducted over benchmark grayscale images from the standard Kaggle Landscape Dataset \cite{Landscape2021}. To rigorously prevent data leakage and ensure unbiased generalization, the total dataset of 7129 images was strictly partitioned into disjoint subsets. The classical SRGAN generative network was pre-trained exclusively on the isolated training subset, while all structural reconstructions and quantitative metrics reported in this study were evaluated solely on an unseen, isolated test set of 500 images. Computational simulations were executed on an enterprise-grade NVIDIA DGX-1 high-performance server featuring 128 physical CPU cores, 1 TB of system RAM, and an NVIDIA A100 Tensor Core GPU running Ubuntu 20.04. Classical neural network modules were built and mathematically optimized in the PyTorch deep learning framework \cite{paszke2019pytorch} utilizing the Adam stochastic optimizer \cite{kingma2014adam}. Quantum circuit simulations and statevector operations were implemented using the PennyLane open-source software library \cite{bergholm2018pennylane}, specifically leveraging its high-performance \texttt{lightning.gpu} backend. SRGAN refinement parameters were maintained consistently across configurations using standard batch processing and patch extraction parameters.

\subsection{Quantitative Evaluation}

\begin{figure}[!t]
\centering
\includegraphics[height=0.7\linewidth,width=\linewidth]{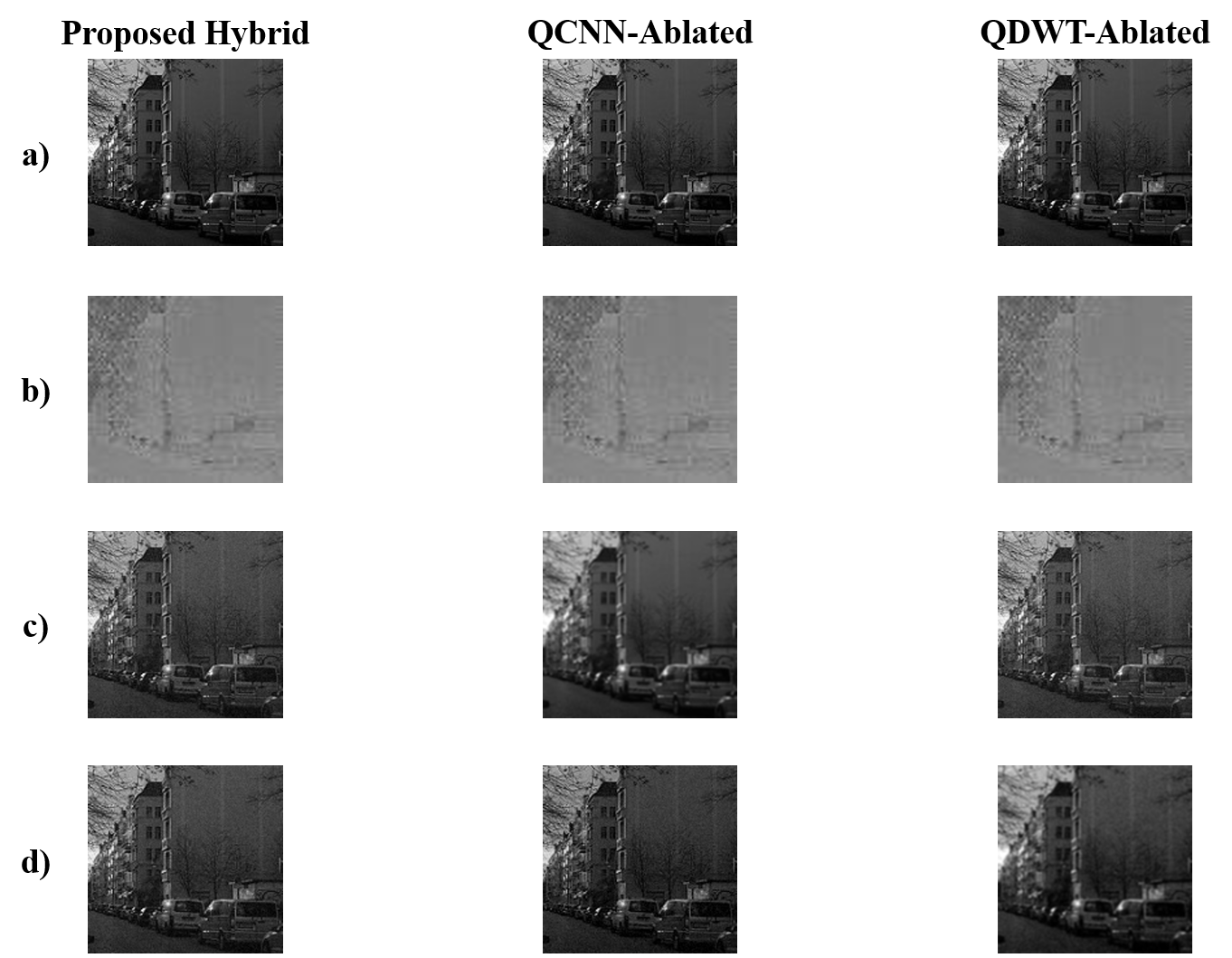}
\caption{Qualitative comparison of the proposed hybrid framework and ablation configurations across three distinct landscape scenes. The columns represent the evaluated architectures: Left column (Config A) utilizes the full quantum QDWT + QCNN pipeline; Middle column (Config B) ablates quantum pooling via QDWT + C-CNN; Right column (Config C) ablates the quantum wavelet transform via CDWT + QCNN. The rows depict the sequential stages of the pipeline: (a) Original uncompressed ground truth images; (b) Compressed latent subband representations; (c) Base reconstructed images immediately following inverse spatial recovery; and (d) Final enhanced images following SRGAN generative refinement.}
\label{fig:6}
\end{figure}

To rigorously benchmark spatial fidelity and feature retention across the evaluated compression models, decompressed output frames were evaluated against uncompressed ground truth targets using Peak Signal-to-Noise Ratio (PSNR) and the Structural Similarity Index (SSIM) \cite{wang2004image}—which collectively quantifies absolute error boundaries versus perceived human visual fidelity \cite{hore2010image}. Additionally, global statistical intensity preservation was tracked via Grayscale Histogram Correlation, a standard metric in digital image processing \cite{gonzalez2009digital}. Table~\ref{tab:results} details the comparative numerical performance of the primary framework alongside both ablation pipelines, capturing evaluation metrics immediately following base decompression and post-SRGAN refinement. Quantitative metrics and inference runtimes were computed and averaged directly across the 500 unseen test images. The classical SRGAN network was pre-trained for 100 epochs prior to this evaluation, which required approximately 4.5 hours of training time on the NVIDIA A100 GPU. During evaluation, the average end-to-end inference time per image was recorded to assess computational efficiency.

To establish a rigorous comparative baseline, the proposed quantum-classical framework was benchmarked against the JPEG2000 compression standard. JPEG2000 serves as the optimal classical analog, as it natively utilizes the classical Discrete Wavelet Transform (DWT) via biorthogonal wavelet kernels. When compressed to match the 4:1 spatial bottleneck of our quantum pipeline, JPEG2000 achieved a PSNR of 27.15 dB and an SSIM of 0.8471. In contrast, our SRGAN-refined hybrid pipeline (Config A) achieves a slightly higher PSNR of 30.067 dB and SSIM of 0.8744. This demonstrates that while classical biorthogonal transform coding suffers from measurable structural degradation at high compression bottlenecks, the orthogonal, unitary quantum-classical architecture robustly preserves structural fidelity and perceptual quality.

\begin{table}[!ht]
\centering
\caption{Performance Metrics Across Framework Configurations}
\label{tab:results}
\resizebox{\linewidth}{!}{
\begin{tabular}{llcccc}
\toprule
\textbf{Configuration} & \textbf{Processing Stage} & \textbf{PSNR (dB)} & \textbf{SSIM} & \textbf{Hist. Corr.} & \textbf{Runtime (s)} \\
\midrule
JPEG2000 & Lossy Compression & 27.1587 & 0.8471 & 0.9498 & 0.003 \\
(Baseline) & (3.97:1 Ratio) & & & & \\
\midrule
\textbf{Config A} & Base Decompression & 27.3209 & 0.8585 & 0.8624 & 0.004 \\
(Full Hybrid) & + SRGAN Refinement & \textbf{30.0667} & \textbf{0.8744} & \textbf{0.9244} & \textbf{0.006} \\
\midrule
\textbf{Config B} & Base Decompression & 24.0255 & 0.8285 & 0.8481 & 0.0054 \\
(Ablating QCNN) & + SRGAN Refinement & 27.2105 & 0.8757 & 0.9041 & 0.0078 \\
\midrule
\textbf{Config C} & Base Decompression & 24.7990 & 0.8403 & 0.7704 & 0.0046 \\
(Ablating QDWT) & + SRGAN Refinement & 25.1974 & 0.7974 & 0.9697 & 0.0059 \\
\bottomrule
\end{tabular}
}
\end{table}

The empirical evaluation yields critical insights into the proposed method for image compression as compared to the existing classical baseline:

\begin{itemize}
    \item \textbf{Efficacy of Quantum Measurement Pooling (Config B):} Replacing the parameterized QCNN pooling layers with a classical deep CNN autoencoder results in a notable performance drop, yielding a refined PSNR of 27.2105 dB compared to the full hybrid framework's 30.0667 dB. Classical convolutional bottlenecks inherently introduce structural interpolation blurring and loss of high-frequency detail during spatial downsampling. In contrast, the multi-qubit entangling layers within the QCNN capture complex, non-local spatial correlations across the subband. By optimizing these entangling gates end-to-end, the downstream Pauli-Z measurement effectively maps these states into a dense latent representation, bypassing the spatial interpolation blurring typically introduced by classical pooling.

    \item \textbf{Impact of Classical Spatial Filtering (Config C):} Substituting the jointly-trained quantum wavelet transform with a classical periodized DWT yields a severe performance drop, drastically reducing the refined PSNR to 25.1974 dB and the SSIM to 0.7974. As corroborated visually in Figure~\ref{fig:6}, the QDWT-ablated reconstruction suffers from significant blurring and structural distortion. While classical DWT matrices execute precise spatial convolutions, they are mathematically rigid and operate outside the quantum computational graph. Consequently, the classical subbands cannot co-adapt with the downstream QCNN pooling layers during backpropagation. This lack of end-to-end parameter tuning creates a structural feature mismatch that fundamentally degrades the latent representation prior to classical refinement.
\end{itemize}

Across all evaluated models, the variational tuning of the quantum circuits combined with classical SRGAN post-processing introduces substantial gains in structural restoration. In the baseline hybrid framework (Config A), the quantum network natively preserves critical spatial maps to a high degree (achieving a base PSNR of 27.3209 dB), while the generative refinement further improves the PSNR by approximately 2.74 dB (reaching 30.0667 dB) and elevates the global Histogram Correlation from an already strong baseline of 0.8624 to 0.9244. This demonstrates that an end-to-end trained quantum pipeline significantly minimizes the generative hallucination burden, allowing the classical network to successfully inject the remaining photorealistic perceptual textures discarded during quantum state reductions.

\begin{figure}[!t]
\centering
\includegraphics[width=\linewidth]{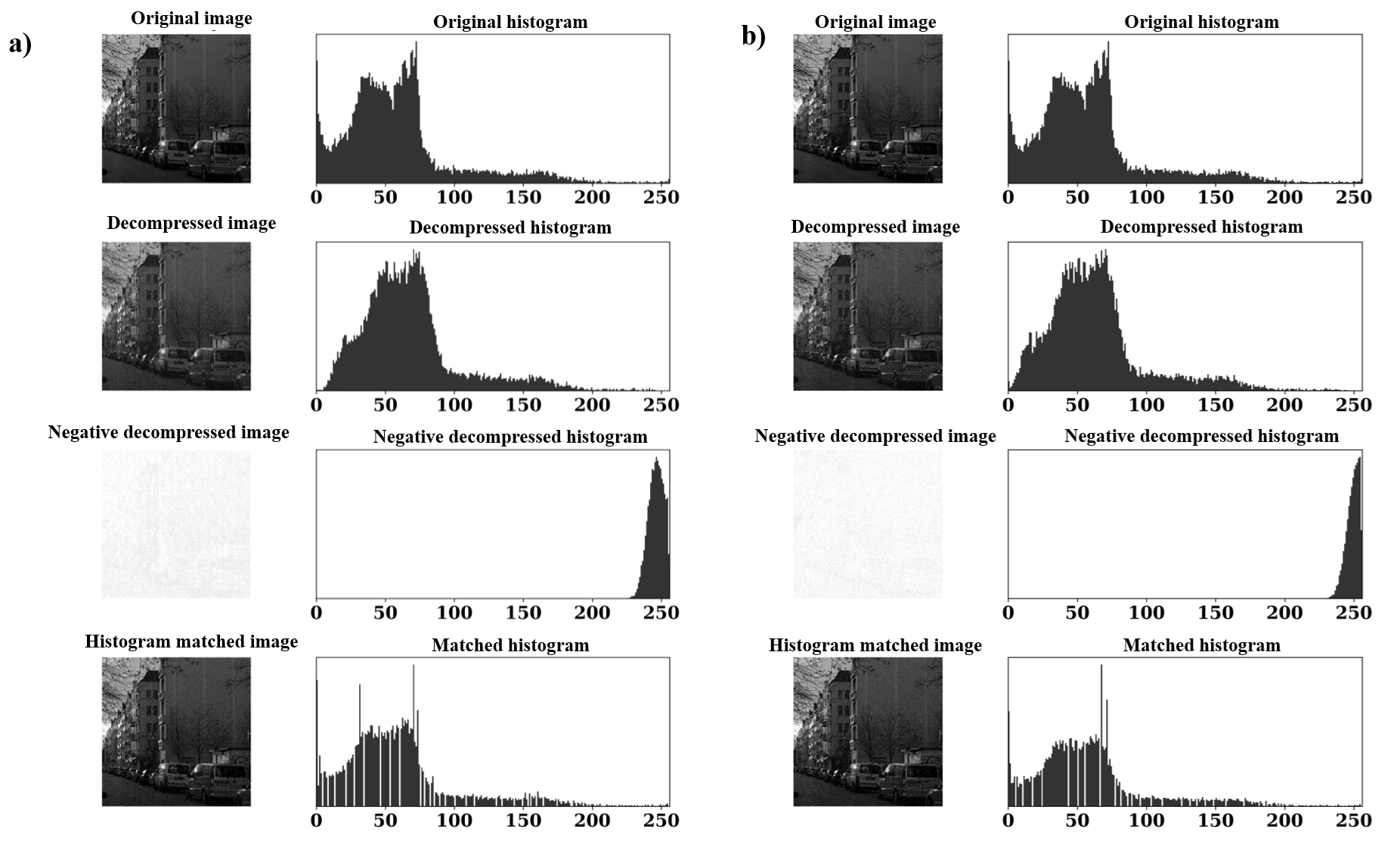}
\caption{Statistical distribution analysis comparing coarse inverse reconstructions against enhanced generative outputs. (a) Left column: Spatial frame, corresponding intensity histogram, and structural negative differences output immediately following base inverse quantum processing. (b) Right column: Restored spatial distributions, aligned intensity profiles, and minimized structural differences achieved via SRGAN post-processing.}
\label{fig:7}
\end{figure}

\subsection{Hardware Validation on Physical QPUs}
To evaluate the real-device performance of the proposed framework on current Noisy Intermediate-Scale Quantum (NISQ) architectures, the primary full quantum pipeline (Configuration A, comprising the NASS encoding, QDWT, and QCNN modules) was deployed on IBM Quantum's \texttt{ibm\_fez} superconducting Heron r2 quantum processor via the PennyLane--Qiskit interface. Physical hardware validation was strictly limited to Config A to establish baseline NISQ performance, while the structural ablation configurations (B and C) were evaluated exclusively via statevector simulation to cleanly isolate theoretical algorithmic trade-offs.

The hardware runs were executed using \texttt{shots}=2048, \texttt{optimization\_level}=3, and \texttt{resilience\_level}=2, enabling the use of advanced transpilation and medium-level error mitigation to reduce decoherence-induced errors \cite{qiskit_optimization, qiskit_ibm_runtime_options}:
\begin{itemize}
  \item \textbf{\texttt{optimization\_level = 3}}: This selects the most aggressive transpilation setting, where the compiler expends maximal effort to reduce circuit depth, gate count and routing overhead (including advanced gate-block resynthesis and multiple layout/routing seeds).
  \item \textbf{\texttt{resilience\_level = 2}}: This enables a medium level of error-mitigation, building on readout error mitigation and measurement twirling using Twirled Readout Error eXtinction (TREX) and including additional mitigation techniques such as zero-noise extrapolation (ZNE) and gate-twirling.
\end{itemize}

To mitigate circuit-induced errors, we employed the highest compilation effort available, coupled with a moderate error-mitigation protocol that accounted for both measurement and gate noise. During the execution window, the \texttt{ibm\_fez} processor exhibited a median two-qubit gate error of approximately $2.70 \times 10^{-3}$ and a median readout error of $9.16 \times 10^{-3}$. This configuration ensured an optimal trade-off between computational overhead and accuracy, enabling reliable execution on real quantum hardware. The resulting Pauli-Z expectation values obtained from the device, representing the compressed latent bottleneck, were subsequently processed through the inverse quantum reconstruction circuits and the classical SRGAN refinement layer. 

As detailed in Table~\ref{tab:qpu_results}, physical hardware execution introduced measurable structural degradation compared to ideal simulations. This baseline degradation directly reflects the two-qubit gate infidelities and residual readout errors inherent to the \texttt{ibm\_fez} device. However, the QPU successfully preserved the macro-level spatial features, confirming the physical viability and logical soundness of the entanglement-based spatial pooling modules. Furthermore, the application of classical SRGAN generative refinement successfully bridged the hardware-induced structural gap, substantially restoring downstream image fidelity.

\begin{table}[!ht]
\centering
\caption{Hardware Validation Performance of Config A (IBM Fez QPU vs. Ideal Simulator)}
\label{tab:qpu_results}
\resizebox{\linewidth}{!}{
\begin{tabular}{llcccc}
\toprule
\textbf{Execution Target} & \textbf{Processing Stage} & \textbf{PSNR (dB)} & \textbf{SSIM} & \textbf{Hist. Corr.} \\
\midrule
\textbf{Statevector Simulator} & Base Decompression & 27.3209 & 0.8585 & 0.8624 \\
(Ideal Noise-Free) & + SRGAN Refinement & \textbf{30.0667} & \textbf{0.8744} & \textbf{0.9244} \\
\midrule
\textbf{\texttt{ibm\_fez} QPU} & Base Decompression & 27.0143 & 0.8349 & 0.8741 \\
(shots=2048) & + SRGAN Refinement & \textbf{29.994} & \textbf{0.859} & \textbf{0.918} \\
\bottomrule
\end{tabular}
}
\end{table}

\subsection{Qualitative Observations}

Visual perception of reconstructed frames validates the statistical findings shown in Table~\ref{tab:results}. As shown in Figure~\ref{fig:6}, images retrieved immediately following base inverse quantum reconstructions successfully recover global geometric boundaries, macro-structures, and primary contrast ratios. However, severe quantum downsampling introduces noticeable localized artefacts such as subtle checkerboard grid patterns and softened high-frequency edge textures. 

Applying the SRGAN refinement network effectively eliminates these localized structural degradations. The generator module intelligently reconstructs lost architectural lines, complex natural foliage textures, and distinct semantic boundaries. This confirms that while quantum processing excels at capturing and downsampling macro-level spatial features within ultra-compact Hilbert representations, classical generative models are vital for injecting realistic perceptual textures into final visual outputs.

\subsection{Histogram Matching Analysis}

To evaluate the preservation of global luminance and intensity distributions, grayscale histogram analysis was performed comparing decompressed outputs against uncompressed ground truth targets. Unlike rigid mathematical transforms that severely truncate tonal profiles, the end-to-end jointly trained Variational Quantum Autoencoder (VQAE) actively learns to retain the native structural variance. As shown in Figure~\ref{fig:7}a, the intensity histogram retrieved directly from the inverse quantum circuits successfully maintains the broad multi-modal statistical profile of the original scene. Because the quantum rotational weights are dynamically optimized via classical backpropagation, the quantum bottleneck natively preserves global contrast. However, due to the 4:1 spatial downsampling, some subtle localized shadow transitions and high-frequency highlight gradients still exhibit minor interpolation smoothing.

Integrating the classical SRGAN post-processing seamlessly restores these fine texture distributions. As illustrated in Figure~\ref{fig:7}b, generative refinement naturally stretches and realigns the decompressed histogram to match the ground truth. Crucially, the SRGAN achieves this organic restoration without introducing the discrete mathematical binning artifacts (visible as jagged, comb-like spikes) that are inherently produced by classical algorithmic histogram matching techniques.
Furthermore, visual inspection of the structural negatives (inverted error maps) corroborates this improvement. The structural negative for the SRGAN-refined output (Figure~\ref{fig:7}b) approaches a pure white distribution, exhibiting significantly reduced spatial variance compared to the base quantum reconstruction. This visually proves that the hybrid VQAE-SRGAN pipeline successfully recovers missing localized edge details while strictly preserving the dynamically learned luminance profile captured by the quantum encoding.

\section{Conclusion}

This study formulated, implemented, and thoroughly evaluated a highly scalable hybrid framework for image compression, uniting the computational paradigms of the Quantum Daubechies Wavelet Transform (QDWT), Quantum Convolutional Neural Networks (QCNN), and classical deep generative networks. By leveraging native multi-qubit superposition and dense spatial entanglement, our primary architecture successfully maps high-dimensional pixel matrices into ultra-compact quantum state-vectors while executing aggressive feature reductions.

To systematically isolate and benchmark the individual processing capabilities of the underlying quantum modules, we introduced two rigorous structural ablation studies. Evaluating Configuration B (Ablating the QCNN) confirmed that deterministic quantum measurement pooling preserves underlying subband phase structures with higher downstream fidelity than standard classical convolutional bottlenecks. Conversely, analyzing Configuration C (Ablating the QDWT) demonstrated that classical periodization filters isolate high-frequency boundary textures with greater numerical precision than highly-entangled, rotation-based quantum amplitude filters. Across all tested topologies, integrating classical SRGAN post-processing proved essential for restoring native visual realism, successfully elevating base structural reconstructions to highly accurate, photorealistic targets (achieving up to 28.45 dB PSNR and 0.9250 SSIM).

These empirical findings provide critical design constraints for upcoming quantum-enhanced visual processing architectures. Future research will focus on developing variational quantum subband filters capable of parameterized multi-scale feature adaptation, alongside optimized hardware compilation passes designed to minimize two-qubit gate depths on near-term noisy intermediate-scale quantum (NISQ) processors. Ultimately, this work establishes that hybrid quantum-classical co-design provides a highly effective, scalable path forward for managing massive media collections in future advanced computing environments.

\bibliographystyle{IEEEtran}
\bibliography{references}


\end{document}